\documentclass{jpp}
\usepackage{graphicx}
\usepackage{tensor}
\usepackage{bm}
\usepackage{enumitem}
\usepackage{stackrel}

\usepackage{tikz}

\newlist{myitemize}{itemize}{3}
\setlist[myitemize,1]{label=\textbullet,leftmargin=2em,rightmargin=2em,itemindent=0pt,labelsep=5pt,labelwidth=2em}
\setlist[myitemize,2]{label=$\rightarrow$,leftmargin=1em}
\setlist[myitemize,3]{label=$\diamond$}

\newlist{myenumerate}{enumerate}{1}
\setlist[myenumerate,1]{leftmargin=3em,rightmargin=3em,itemindent=0pt,labelsep=5pt,labelwidth=2em}

\usepackage{xcolor}
\usepackage[utf8]{inputenc}
\usepackage[T1]{fontenc}
\usepackage{amsmath}

\renewcommand{\tt}{\tilde{\theta}}							

\makeatletter
\newcommand{\vast}{\bBigg@{3}}
\newcommand{\Vast}{\bBigg@{4}}
\makeatother

\title{Intermittent, Reflection-Driven, Strong Imbalanced MHD Turbulence}

\author{B.\ D.\ G.\
  Chandran\aff{1}\corresp{\email{benjamin.chandran@unh.edu}},
N. Sioulas\aff{2}, S. Bale\aff{2}, T. Bowen\aff{2},
  V. David\aff{1}, 
  R. Meyrand\aff{1}, and
  E. Yerger\aff{1} }
\affiliation{\aff{1}Space Science Center and Department of Physics
    and Astronomy,  University of New Hampshire, Durham, NH 03824
\aff{2} Space Sciences Laboratory and Department of Physics,
University of California, Berkeley, CA 94720
}

\begin{document}

\maketitle

\begin{abstract}
  We develop a phenomenological model of strong imbalanced
  magnetohydrodynamic (MHD) turbulence that accounts for intermittency
  and the reflection of Alfv\'en waves  by
    spatial variations in the Alfv\'en speed. Our model predicts the
  slopes of the inertial-range Elsasser power spectra, the scaling
  exponents of the higher-order Elsasser structure functions, and the
  way in which the parallel (to the magnetic field)
  length scale of the fluctuations varies with the perpendicular length
  scale. These predictions agree reasonably well with measurements of
  solar-wind turbulence from the Parker Solar Probe (PSP). In contrast
  to previous models of intermittency in balanced MHD turbulence, we
  find that intermittency in reflection-driven MHD turbulence
  increases the parallel wave numbers of the energetically dominant
  fluctuations at small perpendicular length scales.  This, like the
  PSP measurements with which our model agrees, suggests that
  turbulence in the solar wind and solar corona may lead to more ion
  cyclotron heating than previously realized.
\end{abstract}

\vspace{0.2cm} 
\section{Introduction}
\label{sec:intro}
\vspace{0.2cm}

Magnetohydrodynamic (MHD) turbulence has been an active area of
research for over sixty years. Part of the interest in MHD turbulence
stems from its importance as a model for plasma
turbulence in the solar corona, solar wind, and more distant
astrophysical settings. In this Letter, we focus on non-compressive,
Alfv\'enic fluctuations, for which the energy density of the
magnetic-field fluctuations and velocity fluctuations are
comparable. Such fluctuations are the dominant component of solar-wind
turbulence~\citep{tumarsch95}.

Non-compressive, Alfv\'enic turbulence is conventionally
studied using one of three approximate forms of the MHD equations:
incompressible MHD  \citep{elsasser50}, reduced MHD  
\citep{kadomtsev74,strauss76,schekochihin09}, and a generalization of
the reduced MHD equations that accounts for a radially stratified and
expanding background plasma \citep[see, e.g.,][]{dmitruk03, perez13, vanballegooijen11,vanballegooijen16,vanballegooijen17}. All three approximations are usefully
expressed in terms of the Elsasser variables~\citep{elsasser50},
\begin{equation}
  \bm{z}^\pm = \bm{v} \pm \frac{\bm{B}}{(4\pi \rho)^{1/2}},
  \label{eq:Elsasser_def} 
\end{equation}
where $\bm{v}$, $\bm{B}$, and~$\rho$ are the velocity, magnetic field,
and mass density, respectively. In the presence of a mean magnetic
field $\bm{B}_0$, and when the fluctuations in~$\bm{z}^\pm$ are small
compared to the Alfv\'en
velocity~$\bm{v}_{\rm A} = \bm{B}/\sqrt{4\pi \rho}$, the $\bm{z}^\pm$
fluctuations are Alfv\'en waves that propagate at
velocity~$\mp \bm{v}_{\rm A}$. In Alfv\'enic turbulence in homogeneous
plasmas, only
counter-propagating fluctuations interact nonlinearly
\citep{kraichnan65}. In addition, small-scale Alfv\'enic fluctuations
with correlation length~$\lambda$ propagate along the local background
magnetic field obtained by summing~$\bm{B}_0$ and the magnetic field
associated with all fluctuations with correlation
lengths~$\gg \lambda$ \citep{kraichnan65}.  In contrast to
hydrodynamic turbulence, Alfv\'enic turbulence is intrinsically
anisotropic \citep{montgomery81}, both in the weak-turbulence
regime~\citep{shebalin83,ng96,ng97, galtier00} and strong-turbulence
regime~\citep{goldreich95,cho00}.  In particular, as energy cascades
from large scales to small scales, the small-scale structures that are
produced are elongated along~$\bm{B}$.

Three frontiers in the study of MHD turbulence are imbalance,
intermittency, and inhomogeneity. In imbalanced turbulence, the
fluctuations in one of the Elsasser variables are much larger than the
fluctuations in the other Elsasser variable \citep[see,
e.g.,][]{lithwick07,beresnyak08,chandran08a,perez09a,podesta10,schekochihin22}.
Intermittency refers to the phenomenon in which the majority of the
fluctuation energy at scale~$\lambda$ is concentrated into a fraction
of the volume that decreases as~$\lambda$ decreases \citep[see,
e.g.,][]{frisch96}.  Inhomogeneity, and in particular the variation of
$v_{\rm A}$ with heliocentric distance~$r$, results in the reflection
of Alfv\'en waves \citep{heinemann80,velli93,hollweg07,verdini07}. The
Sun launches only outward-propagating Alfv\'en waves into coronal holes\footnote{Coronal holes are regions of the
  solar corona threaded by ``open magnetic-field lines,'' which
  connect the solar surface to the distant interplanetary medium.}
and the solar wind,
and inhomogeneity leads to the mix of~$\bm{z}^+$ and~$\bm{z}^-$
fluctuations that is needed in order for the fluctuations to interact
\citep{velli89,matthaeus99b,meyrand25}.

In this Letter, we develop a model of Alfv\'enic turbulence
that accounts for all three of these phenomena. We note that
intermittency, which in some ways is the most complicated of the
three, has two relatively simple consequences. First, because
$\bm{z}^+$ fluctuations interact only with~$\bm{z}^-$ fluctuations and
vice versa, and because the
energetically dominant~$\bm{z}^+$ and~$\bm{z}^-$ fluctuations are
confined to largely distinct fractions of the volume, intermittency
makes it harder for a strong~$\bm{z}^+$ fluctuation to~`find' a
strong~$\bm{z}^-$ fluctuation with which to interact. Moreover, since
the volume filling factor~$f_\lambda$ of the energetically
dominant~$\bm{z}^\pm$ fluctuations decreases as~$\lambda$ decreases,
intermittency decelerates the cascade to an increasing degree
as~$\lambda$ decreases, thereby flattening the inertial-range power
spectrum relative to models in which intermittency is
neglected~\citep{maron01}.\footnote{In contrast, in hydrodynamic
  turbulence, eddies interact with themselves, so intermittency acts to
  accelerate the cascade to an increasing degree as the eddy size
  decreases, thereby steepening the inertial-range power spectrum \citep{she94}.}  Second, because $f_\lambda$ decreases
as~$\lambda$ decreases, near the dissipation scale the fluctuations
that dominate the energy have much larger amplitudes than the
root-mean-square fluctuation amplitude computed over the entire
volume, which can enhance the efficiency of certain heating mechanisms
such as stochastic ion heating \citep[see, e.g.,][]{mallet19}.

\section{The inertial range of reflection-driven Alfv\'enic turbulence}
\label{sec:model}

We view Alfv\'enic turbulence as a collection of localized, non-compressive
fluctuations in the Elsasser variables $\bm{z}^\pm$ defined in
(\ref{eq:Elsasser_def}) and take these fluctuations to be
characterized by length scales~$\lambda$ and~$l_\lambda^\pm$
perpendicular and parallel to the magnetic field, respectively. We
define the Elsasser increment
\begin{equation}
  \Delta \bm{z}^\pm_\lambda\left(\bm{x}, \bm{\hat{s}}, t\right)= \bm{z}^\pm \left(\bm{x} + 0.5 \lambda
    \bm{\hat{s}}, t \right)
  - \bm{z}^\pm \left(\bm{x} - 0.5 \lambda
    \bm{\hat{s}}, t \right),
\label{eq:Deltazpm} 
\end{equation}   
where $\bm{\hat{s}}$ is a unit vector perpendicular to~$\bm{B}$, and we take
\begin{equation}
  \delta z^\pm_\lambda (\bm{x},t) = \frac{1}{2\pi} \int_0^{2\pi} {\rm
    d}\theta\,
  \left|\Delta \bm{z}^\pm_\lambda\left(\bm{x}, \bm{\hat{s}}, t\right)\right|
  \label{eq:dzdef}
\end{equation}
to be the characteristic amplitude of the~$\bm{z}^\pm$ structure of
perpendicular scale~$\lambda$ at position~$\bm{x}$ and time~$t$, where
the angle~$\theta$ specifies the direction of~$\bm{\hat{s}}$ within
the plane perpendicular to~$\bm{B}$.

In intermittent turbulence, the tail of the distribution of
  fluctuation amplitudes becomes increasingly prominent as~$\lambda$
  decreases. To account for this, we follow an approach that has been used in two previous
  studies of balanced, intermittent reduced MHD (RMHD) turbulence,
  \cite{chandran15} and \cite{mallet17a}, which we henceforth refer to
  as CSM15 and MS17, respectively.
  In particular, we parameterize the scale-dependence of the
  fluctuation-amplitude
  distribution by treating~$\delta z^\pm_\lambda$ as a random variable given by the
equation
\begin{equation}
  \delta z_\lambda^\pm = \overline{ z}^\pm \beta^q,
  \label{eq:zpm}
\end{equation}
where $\overline{ z}^\pm$ is a scale-independent random number,
$\beta$ is a constant satisfying~$0< \beta <
1$ that we will determine in the analysis to follow,
and~$q$ is a
random integer that is uncorrelated with
$\overline{ z}^\pm$ and that has a Poisson distribution with
scale-dependent mean~$\mu$,\footnote{The probability distribution function
  (PDF) of~$\log_\beta \delta z^\pm_\lambda$ is then the convolution of the PDF of
  $\log_\beta(\overline{ z}^\pm)$ with the discrete Poisson distribution. If
  $\overline{ z}^\pm$ were replaced by a constant, then~$\delta
  z^\pm_\lambda$ would have a log-Poisson distribution.}
\begin{equation}
  P(q) = \frac{e^{-\mu} \mu^{q}}{q!}.
  \label{eq:P}
\end{equation}
For simplicity, we have taken $\mu$ and $\beta$ to be the same 
for $\delta z^+_\lambda$ and~$\delta z^-_\lambda$, but we assume that
\begin{equation}
\left\langle \overline{ z}^+ \right\rangle \gg
\left\langle  \overline{ z}^- \right\rangle ,
\label{eq:imbalance}
\end{equation}
as we consider imbalanced turbulence, where~$\langle \dots \rangle$ denotes an ensemble average.

Given (\ref{eq:zpm}) and~(\ref{eq:P}), the volume filling factor of the most intense fluctuations is~$P(0) =
e^{-\mu}$. As in several previous studies of intermittent MHD
turbulence \citep{grauer94, politano95, chandran15, mallet17a}, we
take this filling factor to be~$\propto
\lambda$ and therefore set
\begin{equation}
  \mu = A + \ln\left(\frac{L_\perp}{\lambda}\right),
  \label{eq:mu} 
\end{equation}
where $A$ is a constant that characterizes the breadth of the
distribution of fluctuation amplitudes at the perpendicular outer
scale~$L_\perp$.  We discuss the possible physical origins of this
scaling in~\S\ref{sec:vff}.

Equation~(\ref{eq:mu}) implies that $\mu \gg 1$ within the inertial range, in
which~$\lambda \ll L_\perp$. It follows from~(\ref{eq:P})
that both the most common value of~$q$ and the median value of~$q$
are~$\simeq \mu$ \citep{choi94}, and hence the most common
fluctuation amplitude at scale~$\lambda$ and the median fluctuation
amplitude at scale~$\lambda$ are approximately
\begin{equation}
  w^\pm_\lambda = \left\langle \overline{ z}^\pm\right\rangle \beta^\mu \propto e^{\mu \ln
    \beta} \propto \lambda^{-\ln \beta}.
  \label{eq:w}
  \end{equation} 
  Fluctuations with $q\ll \mu$ satisfy
  $\delta z^\pm_\lambda \gg w^\pm_\lambda$ and form the tail of the
  distribution, which becomes broader as~$\lambda$ decreases and~$\mu$
  increases.

The functional form of~$P(q)$ 
makes it straightforward to compute 
the $n^{\rm th}$-order structure function of~$\delta z^\pm_\lambda$:
\begin{equation}
  \left\langle \left(\delta z^\pm_\lambda\right)^n \right\rangle=
\left\langle \left( \overline{ z}^\pm\right)^n \right\rangle e^{-\mu}\sum_{q=0}^\infty
\frac{\left(\mu \beta^n\right)^q}{q!} = \left\langle \left( \overline{
    z}^\pm\right)^n \right\rangle e^{-\mu + \mu \beta^n},
  \label{eq:SFn}
\end{equation}
where we have utilized the Taylor expansion $\sum_{q=0}^\infty x^q/q!
= e^x$.
The scaling exponent~$\zeta_n$ of the~$n^{\rm th}$-order structure
function is defined
by the proportionality relation
\begin{equation}
    \left\langle \left(\delta z^\pm_\lambda\right)^n \right\rangle
    \propto \lambda^{\zeta_n}.
    \label{eq:SF2}
\end{equation}
Upon substituting (\ref{eq:mu}) into~(\ref{eq:SFn}), we obtain \citep{chandran15,mallet17a}
\begin{equation}
   \zeta_n = 1 - \beta^n.
  \label{eq:zeta_n}
\end{equation}
As $q$ increases from~0, the summand in
(\ref{eq:SFn}) increases until
$q$  reaches a value $\simeq \mu \beta^n$, after which the summand
decreases with increasing~$q$. The amplitudes of the fluctuations that make the largest
contribution to $\left\langle \left(\delta z^\pm_\lambda\right)^n\right\rangle$
are thus approximately
\begin{equation}
  \delta z^\pm_{(n),\lambda} = \overline{ z}^\pm \beta^{\mu \beta^n}
  \propto \lambda^{-\beta^n \ln \beta}.
  \label{eq:zn}
\end{equation}
It follows from~(\ref{eq:w}) and~(\ref{eq:zn})  that
  $w^\pm_{\lambda} = \delta z^\pm_{(0),\lambda} < \delta z^\pm_{(1),\lambda} <
  \delta z^\pm_{(2), \lambda} < 
  \dots$. As $n$ increases, $\mu \beta^n$ decreases, and~$\delta
  z^\pm_{(n),\lambda}$ characterizes the amplitudes of fluctuations
  that are farther out in the tail of the
  distribution.

We restrict our attention to the strong-turbulence regime, in which
the time $\tau_{\rm nl, \lambda}^-\sim \lambda/\delta z^+_\lambda$ required for
a $\delta z^+_\lambda$ fluctuation to shear a $\delta z^-_\lambda$
fluctuation is comparable to or smaller than the
time~$\tau_{\rm lin, \lambda}^- \sim l_\lambda^+/(2v_{\rm A})$ required for a
point moving with the $\delta z^-_\lambda$ fluctuation at
velocity~$ \bm{v}_{\rm A}$ to pass through
the (counter-propagating)~$\delta z^+_\lambda$ fluctuation. In other
words, we assume that
\begin{equation}
  \chi^+_\lambda \equiv \frac{\delta z^+_\lambda l_\lambda^+}{\lambda
    v_{\rm A}}\gtrsim 1.
  \label{eq:chi_ineq}
\end{equation}
In this limit, each cross section (in the plane perpendicular
to~$\bm{B}$) of a propagating $\delta z^-_\lambda$ structure is
strongly deformed after a time~$\sim \lambda/\delta z^+_\lambda$,
causing the parallel correlation length of the $\delta z^-_\lambda$ fluctuation to
satisfy \citep{lithwick07}
\begin{equation}
  l^-_\lambda \sim \frac{v_{\rm A} \lambda}{ \delta z^+_\lambda}.
  \label{eq:lminus} 
\end{equation}
Two locations within a $\delta z^+_\lambda$
structure separated by a distance~$l^-_\lambda$ along the magnetic
field
are deformed by~$\delta z^-_\lambda$ structures in uncorrelated ways,
and hence~\citep{lithwick07}
\begin{equation}
  l^+_\lambda \simeq l^-_\lambda .
  \label{eq:l_lambda}
  \end{equation} 
Henceforth (except in Section~\ref{sec:CSM15}), we will  drop the distinction between~$l^+_\lambda$
and~$l^-_\lambda$ and use simply~$l_\lambda$. Equations~(\ref{eq:lminus})
and (\ref{eq:l_lambda}) imply that the inequality in
(\ref{eq:chi_ineq})  is replaced by the critical-balance relation
\begin{equation}
  \chi^+_\lambda \sim 1,
  \label{eq:chi_1}
\end{equation}
which states that $\tau^-_{\rm nl,\lambda} \sim \tau_{\rm lin,\lambda}^-$ at all
scales \citep{goldreich95,lithwick07} and within structures of
differing amplitudes at the same scale~\citep{mallet15}.

We define
\begin{equation}
  \epsilon_\lambda^+= \frac{\left(\delta
      z^+_\lambda\right)^2}{\tau_{\rm nl, \lambda}^+} ,
  \label{eq:epslambda}
\end{equation}
which is the rate at which a localized $\delta z^+_\lambda$ fluctuation's
energy cascades to smaller scales, where~$\tau_{\rm nl, \lambda}^+$ is
the energy cascade time scale of the~$\delta z^+_\lambda$ fluctuation.
In reflection-driven MHD turbulence in the solar wind, the
fluctuations propagating towards the Sun in the plasma frame (the
$\bm{z}^-$ fluctuations) are produced by the reflection of the
outward-propagating $\bm{z}^+$ fluctuations
\citep{heinemann80,velli93,hollweg07}. The $\bm{z}^-$ fluctuations are
also subsequently sheared and cascaded by nonlinear interactions with
$\bm{z}^+$ fluctuations \citep{velli89}. As a consequence, in a
reference frame that propagates with the $\bm{z}^+$ fluctuations away
from the Sun, the $\bm{z}^-$ fluctuations at scale~$\lambda$ remain
coherent until the $\bm{z}^+$ fluctuations at scale~$\lambda$ evolve
due to nonlinear interactions (Lithwick et al 2007; see also Section~5
of Chandran \& Perez 2019). \nocite{lithwick07, chandran19} This
`anomalous coherence' implies that the $\bm{z}^+$ fluctuations at
scale~$\lambda$ have a cascade time scale
\begin{equation}
  \tau_{\rm nl,\lambda}^+ \sim \frac{\lambda}{\delta z^-_\lambda},
  \label{eq:taucplus}
\end{equation}
even though this time exceeds the
time~$\sim l_\lambda/2v_{\rm A}$ required for $\bm{z}^+$ and~$\bm{z}^-$
structures at perpendicular scale~$\lambda$ to propagate through each
other at the Alfv\'en speed.

For
values of~$\lambda$ in the inertial range, we assume that the average
cascade flux is independent of~$\lambda$:
\begin{equation}
   \left\langle \epsilon_\lambda^+ \right\rangle \propto \lambda^0.
\label{eq:eps_plus}
\end{equation}
Upon substituting (\ref{eq:epslambda}) and  (\ref{eq:taucplus}) into
(\ref{eq:eps_plus}), we obtain
\begin{equation}
   \left\langle \frac{(\delta z^+_\lambda)^2 \delta
       z^-_\lambda}{\lambda} \right\rangle \propto \lambda^0.
\label{eq:eps_plus2}
\end{equation}
We cannot evaluate the left-hand side of (\ref{eq:eps_plus2}) by
separately averaging  $\left(\delta z^+_\lambda\right)^2$ and
$\delta z^-_\lambda$  using (\ref{eq:zpm}) and~(\ref{eq:P}),
because $\bm{z}^+$ and~$\bm{z}^-$ fluctuations interact with each
other, causing $\delta z^+_\lambda$ and~$\delta z^-_\lambda$ to become
correlated. To model how this correlation affects
$\left\langle \epsilon^+_\lambda \right\rangle$, we assume that
$\left\langle \epsilon^+_\lambda\right\rangle$ is dominated by $\delta
z^+_\lambda$ fluctuations in the
tail of the distribution, which fill
some volume~$\Omega$.  We further assume that, in and around
volume~$\Omega$, the $\bm{z}^-$ fluctuations at scales somewhat larger
than~$\lambda$ are ordinary, with amplitudes comparable to the median
value (\ref{eq:w}) for their scale. These larger-scale fluctuations
provide the $\bm{z}^-$ energy that is fed into fluctuations of
scale~$\lambda$ within volume~$\Omega$, and we make the simplifying
approximation that $\delta z^-_\lambda$ is driven or injected at the same
rate throughout volume~$\Omega$.  We also assume that
nonlinear interactions at scale~$\lambda$ in effect damp
$\delta z^-_\lambda$ on the time scale~$\lambda/\delta z^+_\lambda$,
causing $\delta z^-_\lambda$ to be~$\propto 1/\delta
z^+_\lambda$ in volume~$\Omega$. Choosing the proportionality factor so that
$\delta z^-_\lambda \rightarrow w^-_\lambda$ as
$\delta z^+_\lambda \rightarrow w^+_\lambda$, we find that
\begin{equation}
  \delta z^-_\lambda  \sim \frac{w^-_\lambda w^+_\lambda }{\delta z^+_\lambda}
  \label{eq:dzm}
\end{equation}
in volume~$\Omega$, which corresponds to the tail of the~$\delta
  z^+_\lambda$ distribution.
(Our estimate of~$\delta z^-_\lambda$ in~(\ref{eq:dzm})
  differs from the estimate of~$\delta z^-_\lambda$ in CSM15 for
  reasons that we discuss in \S\ref{sec:CSM15}.)

Upon substituting (\ref{eq:dzm}) into (\ref{eq:eps_plus2}), averaging,
and making use of (\ref{eq:w}), (\ref{eq:SF2}),  and~(\ref{eq:zeta_n}), we obtain 
\begin{equation}
  \left\langle \epsilon_\lambda^+\right\rangle \sim \frac{\left\langle\delta
      z^+_\lambda \right\rangle w^+_\lambda  w^-_\lambda}{\lambda}
  \propto \lambda^{-\beta -2\ln \beta} .
\label{eq:eps_plus_2}
\end{equation}
Equation~(\ref{eq:eps_plus}) then implies that
\begin{equation}
  \beta = - 2 \ln \beta.
  \label{eq:beta}
\end{equation}
The solution to (\ref{eq:beta})  is
\begin{equation}
  \beta = 2 W_0(1/2)  = 0.7035,
  \label{eq:beta2}
\end{equation}
where~$W_0$ is the Lambert~$W$ function.  We note that the $\delta z^+_\lambda$ fluctuations that make the largest
  contribution to~$\left\langle \epsilon^+_\lambda \right\rangle$
  in~(\ref{eq:eps_plus_2}) have amplitudes
  $\simeq \delta z^+_{(1),\lambda} \propto \lambda ^{-\beta \ln \beta}
  = \lambda^{0.247}$, where~$\delta z^+_{(1),\lambda}$ is defined
  in~(\ref{eq:zn}). Deep in the inertial range,
  $\delta z^+_{(1),\lambda} \gg w^+_\lambda \propto \lambda^{-\ln
    \beta} = \lambda^{0.352}$, consistent with our assumption
  that~$\left \langle \epsilon^+_{\lambda}\right\rangle$ is dominated
  by the tail of the~$\delta z^+_\lambda$ distribution.

One might expect that, in reflection-driven MHD turbulence, the
spatial distribution of~$\delta z^-_\lambda$ should mirror the spatial
distribution of~$\delta z^+_\lambda$, so that $\delta z^+_\lambda$
and~$\delta z^-_\lambda$ are highly correlated.  However, the
reflection of inertial-range~$ \bm{z}^+$ fluctuations produces
inertial-range $\bm{z}^-$ fluctuations with a very steep spectrum
\citep{velli89}. As a consequence, the
inertial-range~$\delta z^-_\lambda$ fluctuations are primarily the
result of~$\bm{z}^-$ energy that originates from the reflection of
outer-scale~$\bm{z}^+$ fluctuations and that subsequently 
cascades to
smaller scales. We do not expect the spatial distribution of such
`cascaded' $\delta z^-_\lambda$ fluctuations to approximate the
spatial distribution of $\delta z^+_\lambda$ fluctuations in the
inertial range, and indeed we have argued in~(\ref{eq:dzm}) that
$\delta z^+_\lambda$ and~$\delta z^-_\lambda$ are anti-correlated.

From (\ref{eq:zn}), the fluctuations that make the largest
contribution to the second-order structure function have amplitude
\begin{equation}
  \delta z^\pm_{(2),\lambda} \propto \lambda^{-\beta^2 \ln \beta} =
  \lambda^{0.174},
  \label{eq:zn_2}
\end{equation}
whereas the root mean square fluctuation amplitude scales as
\begin{equation}
\delta z^\pm_{\rm (rms),\lambda} \equiv  \left \langle \left(\delta z^\pm_{\lambda}\right)^2 \right \rangle^{1/2}
  \propto \lambda^{(1-\beta^2)/2} = \lambda^{0.253}.
  \label{eq:zrms}
\end{equation}
The scaling in (\ref{eq:zn_2}) is shallower than in
(\ref{eq:zrms}) because the second-order structure
function is dominated by a fraction $f_\lambda$ of the volume that decreases
as~$\lambda$ decreases, within which the $\bm{z}^\pm$ fluctuations are
unusually strong. This volume filling factor~$f_\lambda$
could be defined in different ways, one being the relation
$ \left[\delta z^\pm_{\rm
    (rms),\lambda}\right]^2 \sim f_\lambda \left[\delta z^\pm
  _{(2),\lambda}\right]^2$. The parallel length scale~$l_\lambda$ that
can be inferred from an analysis of the second-order $\bm{z}^+$ structure
function \citep[see, e.g.,][]{sioulas24} is approximately the value
of~$l_\lambda$ within the fraction~$f_\lambda$ of the volume that
makes the dominant contribution to $\left \langle \left(\delta
    z^+_\lambda \right)^2\right\rangle$, which is approximately
\begin{equation}
  l_{(2),\lambda} \equiv \frac{v_{\rm A} \lambda}{ \delta
    z^+_{(2),\lambda}}  \propto \lambda^{1+\beta^2 \ln \beta} = \lambda^{0.826}.
  \label{eq:l_2}
\end{equation}

To connect our results to the Elsasser power spectra,
we take $E^\pm(f)$ to be the $z^\pm$ power spectrum measured by a
spacecraft in the solar wind, where~$f$ is frequency in the spacecraft
frame. Setting~$f \sim U/\lambda$, where $U$ is
the plasma velocity in the spacecraft frame, and taking $fE^\pm(f)$ to
be 
$\propto \left\langle  \left(\delta z^\pm_\lambda\right)^2\right\rangle$, 
we obtain
\begin{equation}
  E^\pm(f) \propto f^{-1 - \zeta_2} = f^{-1.51}.
  \label{eq:Eplus}
\end{equation}

\section{Comparison with Observations}
\label{sec:obs} 

We compare our model with 
Sioulas et al.'s (2024) \nocite{sioulas24} analysis of
magnetic-field fluctuations during the first perihelion encounter (E1)
of the Parker Solar Probe (PSP), from 1 November, 2018, to 11
November, 2018, when PSP's heliocentric distance~$r$ ranged between
$0.166$ and $0.244$~astronomical units (au).  \cite{sioulas24} analyzed the merged SCaM data
product~\citep{bowen20}, which combines measurements from the
flux-gate magnetometers and search-coil magnetometer of the PSP FIELDS
instrument suite~\citep{bale16}.  \cite{sioulas24} subdivided the data
into 12-hr intervals, with 6~hrs of overlap between consecutive
intervals, and restricted their analysis to intervals in which the
absolute value of the fractional cross helicity
\begin{equation}
 \sigma_{\rm c} = \frac{\left\langle (\delta z^+_\lambda)^2 - (\delta z^-_\lambda)^2 \right\rangle }{
   \left\langle (\delta z^+_\lambda)^2 + (\delta z^-_\lambda)^2 \right\rangle}
 \label{eq:sigmac}
\end{equation}
exceeded~0.75 at~$\lambda = 10^4 d_{\rm i}$, where~$d_{\rm i}$ is
the proton inertial length.
\cite{sioulas24} then computed five-point structure functions of the
magnetic field in a coordinate system in which one axis (the
$l$ axis) is aligned with the local
background magnetic field $\bm{B}_{\rm local}$, a second axis (the
$\xi$ axis) is aligned with the component of the magnetic-field
fluctuation perpendicular to~$\bm{B}_{\rm local}$, and the third
axis (the $\lambda$ axis) is perpendicular to the first two.
Each
increment $\Delta \bm{B}$ computed from the data corresponds to a
spatial separation $\Delta \bm{x}$ parallel to the instantaneous value
of the relative velocity
between the plasma and spacecraft, which in general has a nonzero
component along each of the $l$, $\xi$, and~$\lambda$
directions. \cite{sioulas24} considered three different subsets of
increments, one in which $\Delta \bm{x}$ is within~$5^\circ$ of the
$l$~direction, one in which $\Delta \bm{x}$ is within~$5^\circ$ of
the~$\xi$ direction, and one in which~$\Delta \bm{x}$ is
within~$5^\circ$ of the~$\lambda$ direction. It is this last subset
that we compare with our model results in this section.

To convert time
intervals into spatial intervals, \cite{sioulas24} applied Taylor's frozen-flow
hypothesis based on the average plasma velocity~$U$ in the spacecraft
frame. Plasma
measurements from the SWEAP instrument suite~\citep{kasper16} show
that the average value of $U$ during the selected intervals was
$277 \mbox{ km} \mbox{ s}^{-1}$. Quasi-thermal-noise electron-density
measurements from FIELDS data \citep{moncuquet20} and 
quasi-neutrality imply that the average
value of~$d_{\rm i}$ during the selected intervals was
$16.3 \mbox{ km}$.

\begin{figure}
  \centerline{
  \includegraphics[width=6.5cm]{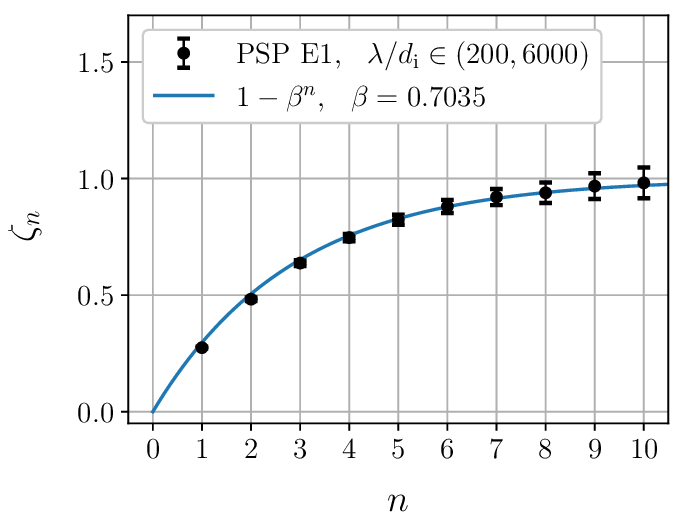}
  \hspace{0.5cm} 
  \includegraphics[width=6.5cm]{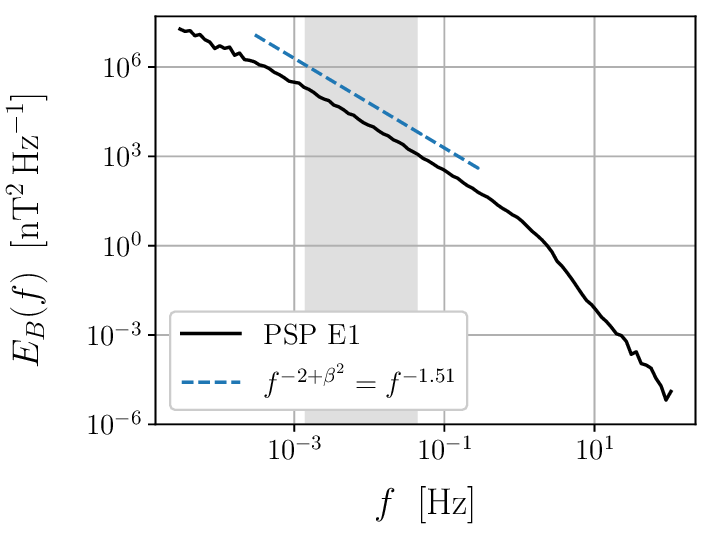}  
}
\caption{{\em Left:} the scaling exponent~$\zeta_n$ of the $n^{\rm
    th}$-order $\bm{z^+}$ structure function from~(\ref{eq:zeta_n})
  and~(\ref{eq:beta2}), and the scaling exponent of the $n^{\rm th}$-order
  magnetic-field structure function obtained by \cite{sioulas24} from
  measurements during PSP's first perihelion encounter.
  {\em Right:} the power spectrum $E_B(f)$ of the magnetic field in
  PSP encounter-1 data as a function of spacecraft-frame frequency~$f$, as
  well as the $\bm{z}^+$ power-spectrum scaling from (\ref{eq:Eplus}). 
The shaded gray rectangle 
shows the frequency interval that corresponds to the scale range~$200
d_{\rm i} < \lambda < 6000 d_{\rm i}$ based on the correspondence
$f = U/(2 \lambda)$ suggested by figure~1 of \cite{huang23}, with $U =
277 \mbox{ km} \mbox{ s}^{-1}$ and~$d_{\rm i} = 16.3 \mbox{ km}$. 
  \label{fig:zeta_n} }
\end{figure}

Within the large-$\sigma_{\rm c}$ intervals analyzed by
\cite{sioulas24}, $\delta z^-_\lambda \ll \delta z^+_\lambda$, and
hence
$\delta z^+_\lambda \simeq 2 \delta B_\lambda/\sqrt{4 \pi \rho}$,
where~$\delta B_\lambda$ is defined by analogy to~(\ref{eq:dzdef}).
We thus expect the structure functions of~$\bm{B}$ in these intervals
to scale with~$\lambda$ in the same way as the $\bm{z}^+$ structures
functions.  In the left panel of figure~\ref{fig:zeta_n}, we plot the
magnetic-field structure-function scaling exponents obtained by
\cite{sioulas24} over the
range~$200 d_{\rm i} < \lambda < 6000 d_{\rm i}$, as well as the
$\bm{z}^+$ structure-function scaling exponents from (\ref{eq:zeta_n})
and~(\ref{eq:beta2}). The error bars on the scaling exponents are
  the errors associated with the power-law fits to the structure
  functions and do not account for errors arising from the sensitivity
  of high-order structure functions to rare events in the tail of
  the distribution \citep[see, e.g.,][]{dudokdewit13}. Further work is
  needed to quantify such errors; here we simply note that the
  observationally inferred $\zeta_n$ values at large~$n$ in
  figure~\ref{fig:zeta_n} should be viewed with caution.

Using the subset of E1 magnetic-field increments obtained by \cite{sioulas24} in which $\Delta \bm{x}$
is within~$5^\circ$ of the~$\lambda$ direction, we compute the
conditional power spectral density of the magnetic field, $E_B(f)$,
obtained via the maximum overlap discrete
wavelet transform~\citep{percival00}, where~$f$ is frequency in the
spacecraft frame. We plot~$E_B(f)$ in the right panel of
figure~\ref{fig:zeta_n}, as well as the Elsasser power-spectrum
scaling from~(\ref{eq:Eplus}).

\begin{figure}
\centerline{
  \includegraphics[width=6.5cm]{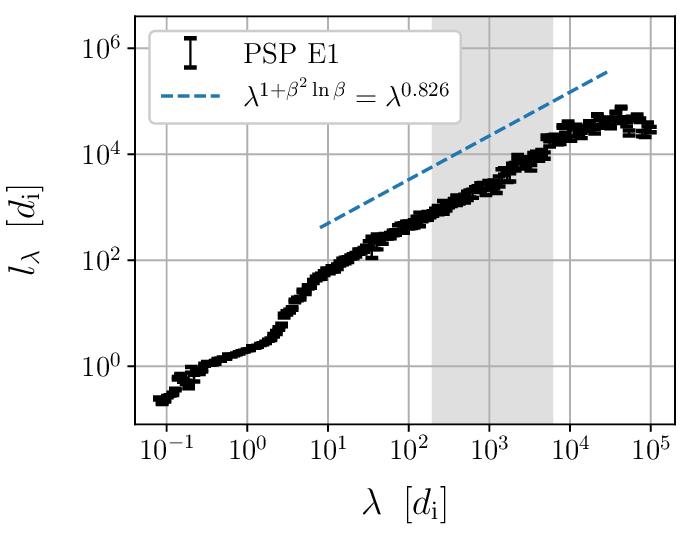}  
}
\caption{
The parallel correlation length $l_\lambda$ inferred from PSP
magnetic-field measurements \citep{sioulas24}, 
and the $l_{(2),\lambda}$ scaling from~(\ref{eq:l_2}).
The shaded rectangle 
corresponds to the scale range that was used to calculate the PSP E1 structure-function
scaling exponents in figure~\ref{fig:zeta_n}.
  \label{fig:l_parallel}  }
\end{figure}

In figure~\ref{fig:l_parallel}  we plot Sioulas et al.'s (2024)
\nocite{sioulas24} result for
the scale-dependent parallel correlation
length $l_\lambda$, which they obtained 
by comparing the parallel second-order structure function~$\mbox{SF}_{2\parallel}(l)$ in which $\Delta
\bm{x}$ is within~$5^\circ$ of the~$l$ direction and the perpendicular second-order structure function
$\mbox{SF}_{2\perp}(\lambda)$ obtained from increments in which~$\Delta
\bm{x}$ is within~$5^\circ$ of the~$\lambda$ direction. They 
defined~$l_\lambda$ to be that value of~$l$ for which
$\mbox{SF}_{2\perp}(\lambda) = \mbox{ SF}_{2\parallel}(l)$.
We also plot in figure~\ref{fig:l_parallel}  the scaling $l_{(2),\lambda} \propto
\lambda^{1+\beta^2 \ln \beta} = \lambda^{0.826}$ from (\ref{eq:l_2})
for the fluctuations that make the largest contribution to the
second-order structure function of~$\bm{z}^+$.

Our model agrees quite well with the power spectrum and $\zeta_n$ values inferred by
\cite{sioulas24} from PSP E1 data for the scale range
$200 d_{\rm i} < \lambda < 6000 d_{\rm i}$. We note, however, that the $\zeta_n$ values that
\cite{sioulas24} found for the scale range
$8 d_{\rm i} < \lambda < 100 d_{\rm i}$ are larger than those shown in
figure~\ref{fig:zeta_n} and do not agree with our model. On the other
hand, the scaling of~$l_{(2), \lambda}$ in~(\ref{eq:l_2}) agrees reasonably well with the~$l_\lambda$
  values inferred by \cite{sioulas24} over the broader scale range $8
  d_{\rm i} < \lambda < 3 \times 10^4 d_{\rm i}$.

\section{Comparison with models of balanced, intermittent, homogeneous RMHD turbulence}
\label{sec:CSM15}

Our model shares several features with the previous models of~CSM15
and~MS17, which
were developed for balanced (i.e., zero-cross-helicity), intermittent,
homogeneous RMHD turbulence. In particular, we adopt the same
procedure in (\ref{eq:zpm}) and~(\ref{eq:P}) for determining the scale
dependence of the fluctuation-amplitude distribution and the same
volume filling factor in~(\ref{eq:P}) and~(\ref{eq:mu}) for the most
intense~$\delta z^+_\lambda$ fluctuations. As a consequence, we obtain
the same formula~(\ref{eq:zeta_n}) as these previous authors for the
scaling exponent~$\zeta_n$ of the $n^{\rm th}$-order structure
function of~$\delta z^+_\lambda$. These shared assumptions receive considerable support from
\cite{sioulas24}'s results on the structure-function scaling exponents
of magnetic fluctuations in the scale range
$200 d_{\rm i} < \lambda < 6000 d_{\rm i}$ in the near-Sun solar wind, which are shown in
figure~\ref{fig:zeta_n}. In particular, \cite{sioulas24}'s finding
that~$\zeta_n$
asymptotes to a constant value at large~$n$ implies that the amplitude
of the most intense fluctuations at scale~$\lambda$ is independent
of~$\lambda$, consistent with~(\ref{eq:zpm}). Moreover, 
  \cite{sioulas24}'s finding
that~$\zeta_n$ asymptotes to~$\simeq 1$ at large~$n$ implies that the
volume filling factor of the most intense fluctuations
is~$\propto \lambda$, consistent with~(\ref{eq:P}) and~(\ref{eq:mu}).

On the other hand, we depart from CSM15 and MS17 by taking
  large-amplitude fluctuations to be tube-like. In contrast, CSM15 and
  MS17 took such fluctuations to be sheet-like, in agreement with
  numerical simulations of homogeneous MHD turbulence
  \citep[e.g.,][]{maron01}.  Our assumption of tube-like fluctuations
  is motivated by \cite{sioulas24}'s results on the three-dimensional
  geometry of the second-order structure function of the magnetic
  field in the scale range $200 d_{\rm i} < \lambda < 6000 d_{\rm i}$
  in PSP data.

  CSM15's assumption that fluctuations in the tail of the
  distribution are
  sheet-like led them to the conclusion that $l^+_\lambda$ is an
  increasing function of~$\delta z^+_\lambda$. To explain this, we
  briefly review CSM15's analysis of the interaction between
  large-amplitude, sheet-like $\bm{z}^+$ structures and much weaker,
  counter-propagating $\bm{z}^-$ fluctuations. CSM15 characterized
  sheet-like $\bm{z}^+$ structures in the tail of the distribution as
  having dimensions $l^+_\lambda$ in the $\bm{B}$ direction,
  $\xi_\lambda$ in the $\Delta \bm{z}^+_\lambda$ direction,
  and~$\lambda$ in the $\bm{B} \times \Delta \bm{z}^+_\lambda$
  direction, with $\xi_\lambda \gg \lambda$ and
  $\xi_\lambda \propto \delta z^\pm_\lambda$. To determine the rate of
  deformation of a sheet-like~$\delta z^+_\lambda$ structure, CSM15
  defined a trial-volume sphere of radius~$\sim \lambda$ located
  somewhere in the middle of the sheet. Noting that the~$\bm{z}^-$
  fluctuations propagating at velocity~$\bm{v}_{\rm A}$ within that
  trial volume have been pre-processed by the part of the $\bm{z^}+$
  sheet upstream of the trial volume, CSM15 defined the ``source
  region'' for the~$\bm{z}^-$ fluctuations in the trial volume to be
  the region upstream of the trial volume closest to the trial volume
within which the~$\bm{z}^-$ fluctuations have not yet
  interacted with the~$\bm{z}^+$~sheet. CSM15 made the approximation
  that the $\bm{z}^-$ fluctuations in this source region are
  median-amplitude $\bm{z}^-$ fluctuations, which are tube-like in the
  CSM15 model. CSM15 used the notation~$l^\ast_\lambda$ to denote the parallel correlation length of such
  median-amplitude fluctuations at perpendicular scale~$\lambda$.
  CSM15 then showed that the $\bm{z}^-$ fluctuations in
  the trial volume that are most effective at shearing the $\bm{z}^+$
  structure are the $\bm{z}^-$ fluctuations that had a
  perpendicular scale $\xi_\lambda$ in the source region. As
  these~$\bm{z}^-$ fluctuations propagate from the source region to
  the trial volume, shearing by the $\bm{z}^+$ sheet reduces their correlation
  length in the~$\bm{B} \times \Delta \bm{z}^+_\lambda$ direction
  from~$\xi_\lambda$ to~$\lambda$ and rotates them into alignment with
  the~$\bm{z}^+$ sheet, so that the angle~$\theta^\pm_\lambda$
  between~$\Delta \bm{z}^+_\lambda$ and~$\Delta z^-_\lambda$ within
  the trial volume is small, which decreases the rate at which the
  $\bm{z}^-$ fluctuations shear the $\bm{z}^+$ structure.

  CSM15 estimated $l^+_\lambda$ using two different arguments. First,
  CSM15 set $l^+_\lambda \sim v_{\rm A} \tau_\lambda^+$, taking the
  parallel correlation length of a~$\bm{z}^+$ fluctuation to be the
  distance it can propagate along the magnetic field before its energy
  cascades to smaller scales. We call this the survival-time argument.
  Second, CSM15 took $l^+_\lambda$ for a~$\bm{z}^+$ structure to be
  the parallel correlation length of the  $\bm{z}^-$
  fluctuations that dominate the shearing of the $\bm{z}^+$~structure
  and evaluating the parallel correlation length of those~$\bm{z}^-$
  fluctuations before they started interacting with the $\bm{z}^+$
  structure -- i.e., in the source region described in the previous
  paragraph. This led to the estimate
  $l^+_\lambda \sim l^\ast_{\xi_\lambda}$.  We call this second
  argument the agent-of-shearing argument.  CSM15 showed that these
  two arguments lead to the same estimate for~$l^+_\lambda$, i.e.,
  that $v_{\rm A} \tau^+_\lambda \sim l^\ast_{\xi_\lambda}$.

  These two arguments were the basis for CSM15's aforementioned
  conclusion that~$l^+_\lambda$ is an increasing function
  of~$\delta z^+_\lambda$.  For the agent-of-shearing argument, this
  is because $\xi_\lambda$ is an increasing function
  of~$\delta z^+_\lambda$, and hence so is~$l^\ast_{\xi_\lambda}$. For
  the survival-time argument, this is because as $\delta z^+_\lambda$
  increases, $\theta^\pm_\lambda$ decreases, the cascade time
  scale~$\tau^+_\lambda$ increases, and the~$\bm{z}^+$ fluctuation
  propagates farther before cascading.\footnote{We note that the
    dominant nonlinear interactions in the CSM15 model exhibit three
    different types of locality. Within the trial volume, the dominant
    nonlinear interactions are between $\bm{z}^+$ and~$\bm{z}^-$
    fluctuations with the same perpendicular scale~$\lambda$.
    When the dimensions of a sheet-like $\bm{z}^+$ fluctuation are
    compared with the dimensions of the tube-like,
    median-amplitude~$\bm{z}^-$ fluctuations of perpendicular
    scale~$\xi_\lambda$ in the source region, the
    perpendicular correlation length of these $\bm{z}^-$ fluctuations
    matches the width~$\xi_\lambda$ of the $\bm{z}^+$ sheet, not the
    thickness~$\lambda$ of the sheet, and the parallel correlation
    length of the~$\bm{z}^-$ fluctuations matches the parallel
    correlation length of the~$\bm{z}^+$ sheet.}

In our model, large-amplitude fluctuations are tube-like and
  therefore their nonlinear interactions are not described by CSM15's
  analysis. In particular, there is no pre-processing of $\bm{z}^-$
  fluctuations as they propagate from a pre-interaction source region
  through an extended sheet-like $\bm{z}^+$ structure, and nonlinear
  interactions do not cause~$\bm{z}^+$ and~$\bm{z}^-$ fluctuations to
  become progressively more aligned as~$\delta z^+_\lambda$ increases.
  We
  estimate~$l^-_\lambda \sim v_{\rm A} \lambda / \delta z^+_\lambda$
  in~(\ref{eq:lminus}) using the survival-time argument. We then use
  the agent-of-shearing argument to estimate~$l^+_\lambda$
  in~(\ref{eq:l_lambda}), setting~$l^+_\lambda \sim l^-_\lambda$. This
  latter estimate makes our survival-time stimate of~$l^-_\lambda$
  consistent with the agent-of-shearing estimate
  of~$l^-_\lambda$. However, our agent-of-shearing estimate
  of~$l^+_\lambda$ differs from the survival-time estimate
  of~$l^+_\lambda$, because
  $v_{\rm A} \tau^+_\lambda \sim v_{\rm A} \lambda / \delta
  z^-_\lambda \gg l^-_\lambda \sim v_{\rm A} \lambda / \delta
  z^+_\lambda$. Thus, although CSM15 did not need to choose between
  the agent-of-shearing argument and the survival-time argument to
  determine~$l^+_\lambda$, we do need to choose, and we choose the
  agent-of-shearing argument.

Whereas CSM15 found that~$l^+_\lambda$ is an increasing
function of~$\delta z^+_\lambda$ in balanced RMHD turbulence, we find
in (\ref{eq:lminus}) and~(\ref{eq:l_lambda}) that $l^+_\lambda$ is a decreasing
function of~$\delta z^+_\lambda$ in imbalanced, reflection-driven,
Alfv\'enic turbulence. 
Because $\delta z^+_{\lambda}$ 
exceeds~$\delta z^+_{\rm rms,\lambda}$ within the small fraction of the
volume that dominates the
second-order~$\bm{z}^+$ structure function (see (\ref{eq:zn_2})
and~(\ref{eq:zrms})), $l^+_\lambda$ is smaller
in this small fraction of the volume than it would be throughout the
plasma as a whole in a turbulence model that neglects
intermittency. Intermittency in reflection-driven
Alfv\'enic turbulence thus increases the effective parallel wave numbers~$k_\parallel
\sim 1/l^+_\lambda$ of the energetically dominant fluctuations at
small~$\lambda$, making these fluctuations more isotropic.

Our assumption that large-amplitude $\bm{z}^+$ fluctuations 
 are tube-like also explains why our
  estimate of~$\delta z^-_\lambda$ in~(\ref{eq:dzm}) differs from
  CSM15's estimate. CSM15 showed that when a
  large-amplitude, sheet-like $\bm{z}^+$ structure interacts with a
  smaller-amplitude $\bm{z}^-$ fluctuation, the amplitude of the
  $\bm{z}^-$ fluctuation is not altered. In contrast, in our analysis,
large-amplitude, tube-like $\bm{z}^+$ fluctuations can alter, and in particular decrease, the amplitudes
  of the $\bm{z}^-$ fluctuations with which they interact.

\subsection{The volume filling factor and geometry of the most intense fluctuations}
\label{sec:vff} 

As mentioned in \S\ref{sec:model} and at the beginning of
\S\ref{sec:CSM15}, we follow previous studies by 
assuming in~(\ref{eq:mu}) that the volume filling factor of the
most intense fluctuations is~$\propto \lambda$ \citep{grauer94,politano95,chandran15,mallet17a}. The usual
justification for this assumption is to posit that the most intense
fluctuations at all~$\lambda$ are associated with the same set of
sheet-like discontinuities. If the two
locations $\bm{x} \pm 0.5 \lambda \bm{\hat{s}}$ that determine the
Elsasser increment $\Delta \bm{z}^\pm_\lambda(\bm{x}, \bm{\hat{s}},t)$
in~(\ref{eq:Deltazpm}) are regarded as a two-point probe, then the
probability that this two-point probe straddles (and therefore
detects) one of these sheet-like discontinuities within some volume~$V$
of turbulent plasma is~$\sim A\lambda/ V$,
where $A$ is the combined area of the sheet-like discontinuities
within that volume.

This line of reasoning is self-consistent within the CMS15 and MS17 models, which
take the large-amplitude fluctuations in the tail of the
distribution to be sheet-like. It becomes problematic, however, in
our model, because we have taken the large-amplitude fluctuations that
dominate the energy cascade to be tube-like, as suggested by
\cite{sioulas24}'s findings regarding the three-dimensional geometry
of the second-order structure function of the magnetic field in the
near-Sun solar wind at $200 d_{\rm i} < \lambda < 6000 d_{\rm i}$. The observed dominance of tube-like fluctuations
in this scale range
might appear to suggest that we should adopt a different scaling for
the volume filling factor of the largest-amplitude
fluctuations. However, the observations reported by \cite{sioulas24} suggest that~$\zeta_n
\rightarrow 1$ as~$n\rightarrow \infty$ within this same range of scales,
an asymptotic behavior that directly implies that the filling factor of the most intense
fluctuations is~$\propto \lambda$, as discussed at the beginning of
\S\ref{sec:CSM15}.

To summarize the previous two paragraphs, the observation
that~$\zeta_n \rightarrow 1$ at large~$n$ suggests that
large-amplitude fluctuations are sheet-like, but the measured
second-order structure function suggests that large-amplitude
fluctuations are tube-like. One possible way to resolve this tension
relates to the fact the second-order structure function is dominated
by fluctuations in the near tail of the PDF with
amplitudes~$\simeq \delta z^+_{(2),\lambda}$, whereas the
filling-factor assumption embedded in~(\ref{eq:mu}) describes
fluctuations in the extreme tail of the PDF with
amplitudes~$\simeq \overline{ z}^+$. These two different parts of the
PDF could arise from different physical
processes. For example, fluctuations in the near tail could be
tube-like because of the (as yet not fully understood) dynamics of the
nonlinear interactions that give rise to the energy cascade in
reflection-driven Alfv\'enic turbulence, while fluctuations in the
extreme tail of the PDF could result from switchbacks ---
sheet-like discontinuities in $\bm{z}^+$ that pervade the near-Sun
solar wind \citep{kasper19,bale19,horbury20}. Switchbacks are likely
produced by the tendency of imbalanced MHD turbulence in compressible
plasmas to evolve towards a state of spherical
polarization~\citep[e.g.,][]{squire20,shoda21,mallet21}.  In this
picture, the energy cascade (see (\ref{eq:eps_plus_2}) and the
discussion following (\ref{eq:beta2})) and second-order
$\delta z^+_\lambda$ structure function are dominated by tube-like
fluctuations in the near tail of the PDF, but the value of~$\zeta_n$
at large~$n$ is controlled by sheet-like switchbacks.  Further work,
however, is needed to determine whether such a hybrid picture of the
PDF is relevant to the solar~wind.

\section{Conclusion}
\label{sec:conclusion} 

In this Letter, we have drawn upon several elements of Lithwick,
Goldreich, \& Sridhar's (2007) \nocite{lithwick07} theory of strong,
imbalanced MHD turbulence to develop a
phenomenological model of intermittent,
reflection-driven Alfv\'enic turbulence. Our treatment of
  intermittency is based upon three principal conjectures.  First, we
  adopt a particular mathematical model, given by~(\ref{eq:zpm})
  and~(\ref{eq:P}), for determining the scale dependence of the PDF of fluctuation amplitudes
  --- the same model that was adopted by~CSM15 and~MS17. Second, we
  assume the same scaling as these authors for the volume filling
  factor of the most intense fluctuations at each scale.  Third, we
  conjecture in~(\ref{eq:dzm}) that, within the small fraction of the
  volume that
  dominates~$\left\langle \epsilon^+_\lambda \right \rangle$ in which
  $\delta z^+_\lambda$ is unusually strong, $\delta z^-_\lambda$
  becomes anticorrelated with~$\delta z^+_\lambda$ because of the
  strong shearing experienced by~$\delta \bm{z}^-$ fluctuations
  where~$\delta z^+_\lambda$ is large.  This third
    conjecture departs from the analysis of CSM15, a contrast that
    results ultimately from our differing assumptions about the
    geometry of the fluctuations in the near tail of the distribution
    that dominate~$\left \langle \epsilon^+_\lambda \right\rangle$
    (see \S\ref{sec:CSM15}). Although our model relies on these
  conjectures, it contains no adjustable fitting parameters. Our
model predicts the scaling of the inertial-range power spectrum, the
structure-function scaling exponent~$\zeta_n$ of
the~$n^{\rm th}$-order structure function for all~$n$, and the 
  scaling of~$l^\pm_\lambda$ with~$\lambda$. These predictions
agree reasonably well with the corresponding scalings inferred by
\cite{sioulas24} from PSP data at
$200 d_{\rm i} < \lambda < 6000 d_{\rm i}$.

Our findings, and the PSP observations with which they agree, have
important implications for the dissipation of reflection-driven
turbulence at small scales. As in previous models of intermittency in
MHD turbulence, we find that the small-scale structures that control
the turbulent heating rate have larger amplitudes than in turbulence
theories that neglect intermittency.
These enhanced amplitudes increase the rate of stochastic ion heating,
as noted in previous studies
\citep[e.g.,][]{chandran10a,xia13, mallet19}. Whereas CSM15 found that intermittency
increases~$l_\lambda$ at small~$\lambda$ within the intense
fluctuations that dominate the energy
in balanced homogeneous MHD turbulence, we find that intermittency in
reflection-driven Alfv\'enic turbulence
has the opposite effect, making the fluctuations that
dominate the energy at small~$\lambda$ more isotropic. This acts to increase the
frequencies of these small-scale fluctuations, possibly to the point
that they can trigger significant ion cyclotron heating, at least in some
regions of the solar corona and solar wind.

Important directions for future research include clarifying and
explaining the amplitude distribution and three-dimensional anisotropy
of fluctuations in the solar wind and in numerical simulations of
reflection-driven Alfv\'enic turbulence and determining the reasons
for the differences between the turbulence scalings seen in PSP data
at $200 d_{\rm i} < \lambda < 6000 d_{\rm i}$ and
$8 d_{\rm i} < \lambda < 100 d_{\rm i}$ \citep{sioulas24}.  Other
useful directions for future research include
using direct numerical simulations and a wider variety of observational data to test
  our results and modeling assumptions and exploring the
consequences of our model for ion cyclotron heating in the solar
corona and solar wind.

\acknowledgements

BC thanks Alfred Mallet and Alex Schekochihin for many valuable
discussions of intermittency in MHD turbulence and Joe Hollweg, Jean
Perez, Jono Squire, and Marco Velli for helpful discussions of
reflection-driven turbulence. The authors thank the anonymous reviewers
for helpful comments and criticisms that led to improvements in the manuscript.
This work was supported in part by NASA grant NNN06AA01C
to the Parker Solar Probe FIELDS Experiment and by NASA grants
80NSSC24K0171
and 80NSSC21K1768.

\bibliography{articles}

\bibliographystyle{jpp}

\end{document}